\documentclass[bibyear]{aa}
\usepackage{graphicx}
\usepackage{txfonts}
\usepackage{amstext}
\usepackage{color}

\begin{document}

   \title{Merging of a CO WD and a He-rich white dwarf to produce a type Ia supernovae}


   \author{D. Liu\inst{1,2,3,4},
          B. Wang\inst{1,2,3,4},
          C. Wu\inst{1,2,3,4}
          \and
          Z. Han\inst{1,2,3,4}
          }
   \institute{Yunnan Observatories, Chinese Academy of Sciences, Kunming 650216, China\\
              \email{liudongdong@ynao.ac.cn; wangbo@ynao.ac.cn}
          \and
          Key Laboratory for the Structure and Evolution of Celestial Objects, Chinese Academy of Sciences, Kunming 650216, China
          \and
          University of Chinese Academy of Sciences, Beijing 100049, China
          \and
          Center for Astronomical Mega-Science, Chinese Academy of Sciences, Beijing, 100012, China}

   \date{}

  \abstract
   {Although type Ia supernovae (SNe Ia) play a key role in astrophysics, the companions of the exploding carbon-oxygen white dwarfs (CO WDs) are still not completely identified. It has been suggested
recently that a He-rich WD (a He WD or a hybrid HeCO WD) that
merges with a CO WD may produce an SN Ia. This theory was based on the double-detonation model, in which the shock compression in the CO core caused by the surface explosion of the He-rich shell might lead to the explosion of the whole CO WD. However, so far, very few binary population synthesis (BPS) studies have
been made on the merger scenario of a CO WD and a He-rich WD in the context of SNe Ia.}
   {We aim to systematically study the Galactic birthrates and delay-time distributions of SNe Ia based on the merger scenario of a CO WD and a He-rich WD.}
   {We performed a series of Monte Carlo BPS simulations to investigate the properties of SNe Ia from the merging of a CO WD and a He-rich WD based on the Hurley rapid binary evolution code. We also considered the influence of different metallicities on the final results.}
   {From our simulations, we found that no more than 15\% of all SNe Ia stem from the merger scenario of a CO WD and a He-rich WD, and
   their delay times range from $\sim$$110\,\rm Myr$ to the Hubble time. This scenario
   mainly contributes to SN Ia explosions with intermediate and long delay times.
   The present work indicates that the merger scenario of a CO WD and a He-rich WD can roughly reproduce the birthrates of SN 1991bg-like events, and cover the range of their delay times. We also found that SN Ia birthrates from this scenario would be higher for the cases with low metallicities.}
   {}

\titlerunning{Merging of a CO WD and a He-rich WD to produce SNe Ia}

\authorrunning{D. Liu et al.}

   \maketitle
%

\section{Introduction} \label{1. Introduction}
Type Ia supernovae (SNe Ia) are defined as SN explosions with no H or He lines in their spectra, but with strong SiII absorption lines near their maximum light (e.g., Filippenko 1997). They are highly luminous events that have great influence on the studies of precision cosmology and galactic chemical evolution (e.g., Matteucci \& Greggio 1986; Riess et al. 1998; Perlmutter et al. 1999). However, the progenitor systems and explosion mechanisms of SNe Ia are still uncertain (Branch et al., 1995; Hillebrandt \& Niemeyer, 2000). It is widely accepted that SNe Ia originate from thermonuclear explosions of carbon-oxygen white dwarfs (CO WDs) in close binary systems (e.g., Hoyle \& Fowler 1960). Nevertheless, the companion stars of exploding WDs are still not absolutely identified (e.g., Podsiadlowski et al. 2008; Howell 2011; Wang \& Han 2012;  Maoz et al. 2014). Up to now, two popular progenitor models of SNe Ia have been proposed, in which the companion could be a non-degenerate star for the single-degenerate model (e.g., Whelan \& Iben 1973; Nomoto et al. 1984) or another He/CO WD for the double-degenerate model (e.g., Iben \& Tutukov 1984; Webbink 1984). For more discussions on the progenitors of SNe Ia, see, for instance, Hachisu et al. (1996), Li \& van den Heuvel (1997), Langer et al. (2000), Nelemans et al. (2001), Han \& Podsiadlowski (2004), Chen et al. (2012), Liu et al. (2012), Wang et al. (2013a), and Zhou et al. (2016).

Alternatively, it has recently been suggested that merger of a CO WD and a He-rich WD may produce SNe Ia through the double-detonation explosion mechanism (e.g., Dan et al. 2012, 2014, 2015; Pakmor et al. 2013; Papish et al. 2015).\footnote{We note that the violent merger of double CO WDs can also produce SNe Ia if the mass ratio of the double WDs is higher than $\sim$$0.8$ (e.g., Pakmor et al 2010, 2011; R\"opke et al. 2012; Ruiter et al. 2013). Liu et al. (2016) recently estimated that the overall SN Ia birthrate from the violent merger scenario accounts for at most 10\% of the inferred observational results in the Galaxy.} This could occur via steady accretion or through a merger. The companion star of the CO WD is usually a He-burning star or a helium-rich WD (e.g., Bildsten et al. 2007; Shen \& Bildsten 2009; Guillochon et al. 2010; Sim et al. 2010; Wang et al. 2013b; Ruiter et al. 2014; Geier et al. 2015). It has been proposed that the detonation of the He-rich shell is triggered via thermal instability if the companion of the CO WD is a He star (e.g., Nomoto 1982), whereas the detonation of the He-rich envelope is ignited dynamically if the companion is a He-rich WD (e.g., Guillochon et al. 2010). The merging of a CO WD and a He-rich WD may produce double-detonation phenomena through the subsequent process. When the merging of a CO WD and a He-rich WD starts, the mass-transfer would be dynamically unstable and the surface of the CO WD is directly impacted by the accretion streams. When the He-rich torus is so large that the accretion streams are deflected and no longer directly impact the surface of the CO WD, Kelvin-Helmholtz instabilities would appear, leading to the detonation of the He-rich envelope (Guillochon et al. 2010). The shock waves into the CO core driven by the first detonation of the He-rich envelope may trigger a second detonation that would destroy the whole CO WD, corresponding to subluminous SNe Ia (e.g., Fink et al. 2010; Sim et al. 2010; Woosley \& Kasen 2011; Pakmor et al. 2013; Dan et al. 2014). If it fails to trigger the second detonation of the CO core, it may resemble an SN .Ia event (e.g., Shen et al. 2010; Waldman et al 2011; Piersanti et al. 2015; Brooks et al. 2015).

In recent years, the merging of a CO WD and a He-rich WD for producing SNe Ia has been studied extensively.
By considering the influence of a thin He shell on the surface of the CO WDs, Pakmor et al. (2013) hypothesized that the merging of double WDs will trigger double detonations, and speculated that the merging of CO WD$+$He/CO WD systems might explain brightness distributions, delay times, and birthrates of SN Ia observations. Dan et al. (2012, 2014) investigated the outcomes of 225 double WD mergers with a wide range of WD masses and chemical compositions. They found that the merging of WD binaries containing a He-rich WD can achieve the conditions to trigger a detonation in the core of CO WD, and obtained the parameter space of CO WD$+$He-rich WD systems that can produce SNe Ia. Dan et al. (2015) recently suggested that the merging of a $0.45\,M_{\odot}$ He-rich WD$+$$0.9\,M_{\odot}$ CO WD system is consistent with the properties of subluminous 1991bg-like events. Moreover, Wang \& Li (2014) found that the yield of $^{\rm 44}\rm{Ti}$ in the Tycho SN is above $10^{\rm -4}\,M_{\odot}$, and suggested that a He WD is needed to produce this much $^{\rm 44}\rm{Ti}$ in the double-degenerate model (see also Troja et al. 2015).

Crocker et al. (2016) recently studied the SN Ia progenitor scenario involving CO WD$+$pure He WD mergers with the binary population synthesis (BPS) method. They suggested that this population can plausibly account for the antimatter source in the Galaxy as the merging of CO WD$+$ pure He WD systems satisfies the age and birthrate requirements of a Galaxy positron source and matches the $^{\rm 44}$Ti yield requirements. Crocker et al. (2016) argued that the merging of CO WD$+$pure He WD systems may produce subluminous 1991bg-like events.

It has been proposed that hybrid HeCO WDs have a CO-rich core surrounded by a He-rich mantle (e.g., Iben \& Tutukov 1985). These WDs are thought to be produced after the He-core-burning stage of hot subdwarfs (e.g., Iben \& Tutukov 1985; Tutukov \& Yungelson 1996; Justham et al. 2011). The hot subdwarfs in binary systems are formed by the stripping of almost all the hydrogen envelope from a red giant star through interaction with a close companion (Han et al. 2002, 2003). Rappaport et al. (2009) recently also presented a binary evolutionary way for the production of
a CO WD$+$He-rich WD system. They found that a $3.4\,M_{\odot}$ star would evolve into a $0.475\,M_{\odot}$ He-core burning subdwarf after the evolution of a common envelope (CE), and would eventually produce a hybrid HeCO WD that comprises a He envelope with about 30\% of its total mass.

By assuming that a detonation would be triggered when the local dynamical timescale is greater than the thermonuclear timescale, Dan et al. (2012) obtained the parameter space of CO WD$+$He-rich WD systems that could form SNe Ia. However, there exists very little corresponding BPS research for the CO WD$+$He-rich WD scenario in context of SNe Ia so far. In this work, we aim to investigate the birthrates and delay times of SNe Ia for the CO WD$+$He-rich WD scenario using a detailed Monte Carlo BPS method. In Sect.\,2 we describe our numerical code for BPS calculations. The results are given in Sect.\,3. A detailed discussion and a summary are provided in Sect.\,4.

\section{Numerical code and methods} \label{2. Methods}
\subsection{Criteria for SN explosions}
Employing the Hurley rapid binary evolution code (Hurley et al. 2000, 2002), we performed a series of Monte Carlo BPS simulations to investigate the properties of SNe Ia based on the CO WD$+$ He-rich WD scenario. Here, we evolved $1\times10^7$ primordial binaries in each set of simulations. In order to examine the influence of metallicity on the birthrates of SNe Ia from the CO WD$+$He-rich WD scenario, we set metallicities Z to be 0.0001, 0.001, 0.02, and 0.03 in our simulations.

In our calculations, we assumed that WDs with masses lower than $0.6\,M_{\odot}$ are He-rich WDs (e.g., Dan et al. 2012, 2014; Yungelson \& Kuranov 2016), which may not be realistic. This assumption aims to give an upper limit of the SN Ia birthrate based on the merger scenario of a CO WD and a He-rich WD. Han et al. (2000) calculated the evolution of low- and intermediate-mass close binaries and described the final fate of these systems. According to their calculations, HeCO WDs with masses higher than $0.55\,M_{\odot}$ have a He mantle larger than $0.05\,M_{\odot}$ (see Table A1 of Han et al. 2000). Alternatively, we also set the upper mass limit of He-rich WDs to be $0.55\,M_{\odot}$ to test its effects on the final results. In our BPS code, a He WD is formed by the complete envelope loss of a first giant branch (FGB) star with $M < M_{\rm HeF}$, while a CO WD is formed by the envelope loss of a thermally pulsing asymptotic giant branch (TPAGB) star with $M < M_{\rm up}$ (see Hurley et al. 2000), where $M$ is the mass of the FGB or AGB star, $M_{\rm HeF}$ is the maximum initial mass of the FGB star that undergoes degenerate He ignition, and $M_{\rm up}$ is the minimum initial mass of the AGB star that undergoes non-degenerate C ignition (see Table 1 of Pols et al. 1998).

In the present work, we assumed that an SN Ia would be formed when the masses of CO WD$+$He-rich WD systems are located in the region of the parameter spaces for producing SN explosions (see Fig.\,2 of Dan et al. 2012). We employed the parameter space obtained through the SPH-smoothed temperature in Dan et al. (2012). In addition, the delay times of SNe Ia are shorter than the Hubble time. The delay time is defined as the time interval from the formation time of the primordial binaries to the merging time of CO WD$+$He-rich WD systems.

\subsection{Evolutionary channels to CO WD$+$He-rich WD}
\begin{figure*}
   \centering
   \includegraphics[width=15cm,angle=0]{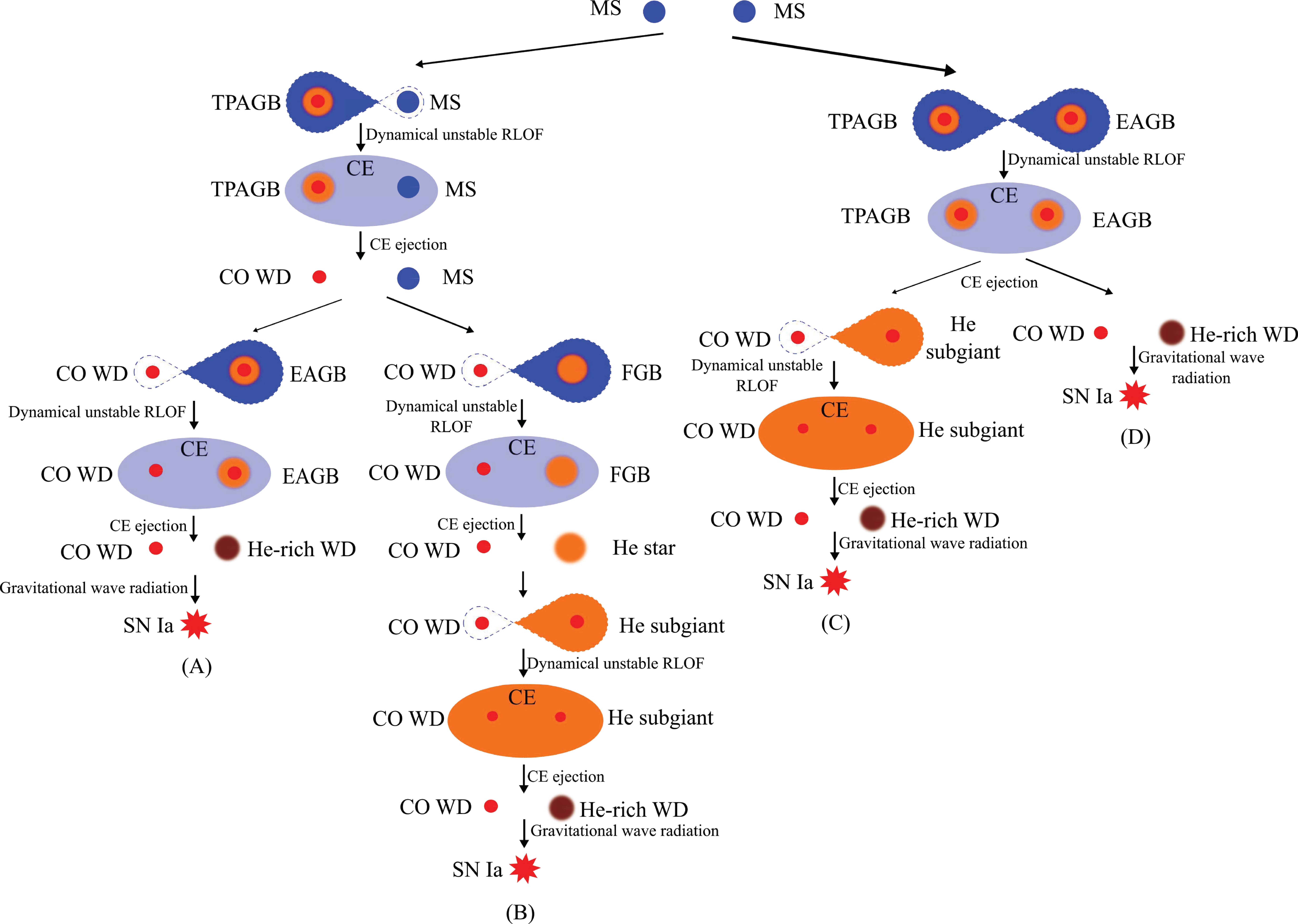}
 \caption{Binary evolutionary channels for producing SNe Ia in the merger scenario of a CO WD and a He-rich WD.}
\end{figure*}

Four formation channels can form CO WD$+$He-rich WD systems and then produce SNe Ia, depending on the evolutionary phase of the primordial secondary after the first ejection of the CE (see Fig.\,1), as described below.

{\em Channel} A: When the primordial primary evolves into the TPAGB stage, it first fills its Roche lobe and starts to transfer material to the primordial secondary. In this case, the mass
transfer is dynamically unstable, resulting in a CE process. After the ejection of the CE, a CO WD$+$MS system is formed, and the secondary continues to evolve. When the secondary evolves into the early AGB (EAGB) stage, it fills its Roche lobe and the mass-transfer is dynamically unstable at this stage, leading to the formation of a CE. If the CE can be ejected, a CO WD$+$He-rich WD system is produced. The CO WD$+$He-rich WD system may eventually merge through gravitational wave radiation and form an SN Ia based on the double-detonation model. For this channel, the parameters of the primordial binaries are in the range of $M_{\rm 1,i}$$\sim$2.6$-$$6.0\, M_{\odot}$, $q=M_{\rm2,i}/M_{\rm1,i}$$\sim$$0.32$$-$0.93, and $P^{\rm i}$$\sim$2200$-$$6300\,\rm days$.

{\em Channel} B: The formation of CO WD$+$MS systems for this channel is similar to the formation in {\em Channel} A. After this, the secondary continues to evolve and fills its Roche lobe at the first giant branch (FGB) stage. In this case, the mass
transfer is dynamically unstable and a CE may be formed. If the CE can be ejected, the secondary is a He Hertzsprung gap (HG) star. The He HG star continues to evolve and will expand quickly, with the result that the He HG star fills its Roche lobe again. In this case, the mass transfer is dynamically unstable, leading to a CE process. If the CE can be ejected, a CO WD$+$He-rich WD system is produced. If the CO WD$+$He-rich WD system can merge through the gravitational wave radiation, an SN Ia is produced eventually. For this channel, the parameters of the primordial binaries are in the range of $M_{\rm 1,i}$$\sim$4.2$-$$6.1\, M_{\odot}$, $q=M_{\rm2,i}/M_{\rm1,i}$$\sim$$0.59$$-$0.94, and $P^{\rm i}$$\sim$3600$-$$6700\,\rm days$.

{\em Channel} C: The primordial primary first fills its Roche lobe when it evolves to the TPAGB stage and the primordial secondary evolves to the EAGB stage. In this case, a CE may be formed as
a result of the dynamically unstable mass transfer. If the CE can be ejected, the primary becomes a CO WD and the secondary a He HG star. Subsequently, the He HG star continues to evolve and expands quickly, leading to the formation of a CE after the He HG star fills its Roche lobe again. If the CE can be ejected, a CO WD$+$He-rich WD system may be formed, and an SN Ia is eventually produced after the merging of the CO WD$+$He-rich WD system. For this channel, the parameters of the primordial binaries are in the range of $M_{\rm 1,i}$$\sim$2.4$-$$6.1\, M_{\odot}$, $q=M_{\rm2,i}/M_{\rm1,i}\ge 0.94,$ and $P^{\rm i}$$\sim$1200$-$$5000\,\rm days$.

{\em Channel} D: Similar to Channel \textup{}B, a CE is formed when the primary is a TPAGB star and the secondary is an EAGB star as a result of the dynamically unstable mass transfer at this stage. If the CE can be ejected, the primary becomes a CO WD and the secondary a He-rich WD, or in other words, a CO WD$+$He-rich WD system is produced. An SN Ia is eventually produced if the CO WD$+$He-rich WD system can merge via the gravitational wave radiation. For this channel, the parameters of the primordial binaries are in the range of $M_{\rm 1,i}$$\sim$2.5$-$$3.2\, M_{\odot}$, $q=M_{\rm2,i}/M_{\rm1,i}$$\sim$1, and $P^{\rm i}$$\sim$230$-$$1500\,\rm days$.

Karakas et al. (2015) have proposed three channels for the production of He-rich WD$+$CO WD mergers to explore the dust production rate in R Coronae Borealis stars. Since the masses of double WDs in Karakas et al. (2015) are significantly lower than those in the present work (the total masses of double WDs in Karakas et al. (2015) are lower than $1.05\, M_{\odot}$, while those in this work are higher than $1.2\, M_{\odot}$), the formation channels between these two works are quite different. In Karakas et al. (2015), the He-rich WDs are produced by the Roche-lobe overflow of subgiant stars that originated from the primordial primaries, and all of their channels only involve one CE event when the secondary is at the AGB stage. In the present work, the He-rich WDs originate from the primordial secondaries that have experienced a CE process, and Channels{\em } A, B, and C involve two CE events.

\subsection{Common-envelope calculations}
When the primordial primary fills its Roche lobe in the binary, the mass ratio ($q'$$=$$M_{\rm donor}/M_{\rm accretor}$, where $M_{\rm donor}$ is the mass of the donor and $M_{\rm accretor}$ is the mass of the accretor) and the evolutionary states of both stars are crucial for the stability of mass transfer. When the mass ratio is greater than a critical ratio, that is, $q'>q_{\rm c}$, the mass transfer is dynamically unstable, leading to the formation of a CE (Paczy\'{n}ski 1976). The critical ratio varies with the evolutionary stage of the donor when it fills its Roch lobe (Hjellming \& Webbink 1987; Webbink 1988; Han et al. 2002; Podsiadlowski et al. 2002). In the present work, we assumed $q_{\rm c}$$=$$4.0$ when the donor is in the HG stage, which is supported by detailed binary calculations (Han et al. 2000). When the donor is an FGB or AGB star, we employed
\begin{equation}
q_{\rm c}=[1.67-x+2(\frac{M_{\rm c1}}{M_{\rm 1}})^{\rm 5}]/2.13
,\end{equation}
where $M_{\rm c1}$, $M_{\rm 1}$ and $x$$=$$d\,lnR_{\rm 1}/d\,lnM_{\rm 1}$ are the core mass of the donor, the donor mass, and the mass-radius exponent of the donor, respectively. When the donor is an MS star, core He-burning star or MS He-naked star, $q_{\rm c}$ was set to be 3. When the donor is an HG-naked He giant star or a giant branch He star, we assumed $q_{\rm c}=0.748$ based on Equation (1) (for details see Hurley et al. 2002).

After a CE has been formed, the primordial secondary and a compact core of the primordial primary are embedded in the CE. In this case, the orbit of the binary shrinks because of the frictional drag within the envelope. During this spiral-in process, a large part of the released orbital energy is injected into the envelope, which may lead to the ejection of the CE (Livio \& Soker 1988).
Nevertheless, it is still an open question how the ejection of the CE is to be calculated (e.g. Ivanova et al. 2013). Here, we used the standard energy prescription described in Webbink (1984) to calculate this process, which is written as
\begin{equation}
\alpha_{\rm CE}(\frac{GM^{\rm f}_{\rm don}M^{\rm f}_{\rm acc}}{2{a_{\rm f}}}-\frac{GM^{\rm i}_{\rm don}M^{\rm i}_{\rm acc}}{2{a_{\rm i}}})= \frac{GM^{\rm i}_{\rm don}M_{\rm env}}{\lambda R_{\rm don}},
\end{equation}
where $M_{\rm don}$, $M_{\rm acc}$, $a$, $M_{\rm env}$ , and $R_{\rm don}$ are the mass of the donor, the mass of the accretor, the orbital separation, the mass of the donor envelope, and the radius of the donor, respectively. The indices i and f represent the values before and after CE process, respectively.
From Equation (2), we can see that there are two alterable parameters in this prescription, which are the stellar structure parameter ($\lambda$) and the CE ejection efficiency ($\alpha_{\rm CE}$) (e.g., Wang et al. 2009).

The stellar structure parameter $\lambda$ was set to be constant ($\sim$0.5; e.g., de Kool 1990; Dewi \& Tauris 2000; Tauris \& Dewi 2001; Hurley et al. 2002), which might be far from the reality (e.g., Xu \& Li 2010). In this work, we employed a variable $\lambda$ that is implicitly included in a fitting formulae of envelope binding energy ($E_{\rm bind}$) (Loveridge et al. 2011; Zuo \& Li 2014). $E_{\rm bind}$ is computed by integrating the gravitational and internal energies:
\begin{equation}
E_{\rm bind}=\int^{M_{\rm s}}_{M_{\rm c}}E_{\rm in}{\rm d}m-\int^{M_{\rm s}}_{M_{\rm c}}(\frac{Gm}{r(m)}){\rm d}m
,\end{equation}
where $M_{\rm c}$, $M_{\rm s}$, $m$ and $E_{\rm in}$ are the mass from the center to the core-envelope boundary, the mass from the center to the surface of the envelope, the mass coordinate, and the internal energy per unit mass. Here, the $E_{\rm in}$ contains the radiation energy and the thermal energy of the gas, but no recombination energy (for details, see van der Sluys et al. 2006).
After the internal energy of the primary envelope is taken into account, the CE ejection efficiency $\alpha_{\rm CE}$ was restricted to be less than 1 (e.g., Davis et al. 2012). Thus, we set the CE ejection efficiency $\alpha_{\rm CE}=0.3$, 0.5, and 1.0 to examine its influence on the final results.

\subsection{Basic assumptions in BPS simulations}
In our BPS calculations, there are some basic assumptions as follows: (1) All stars are treated as members of binary systems with circular orbits. (2) The initial mass function of Miller \& Scalo (1979) is employed for the mass distribution of primordial primaries. (3) In order to generate the mass distribution of the primordial secondaries, the initial mass ratio is assumed to be constant, that is., $n(q)=1$. (4) The distribution of initial orbital separations is assumed to fall off smoothly for close binaries and be constant in $\log\,a$ for wide binaries, where $a$ is orbital separation (e.g., Eggleton et al. 1989). This distribution indicates that approximately half of the stellar systems have orbital periods shorter than $100\,\rm yr$, and that the number of wide binary systems is equal in each logarithmic interval. (5) The star formation rate is assumed to be constant ($5\,M_{\odot}\,\rm yr^{\rm -1}$) over the past $14\,\rm Gyr$ to provide an approximate description of spiral galaxies, or alternatively, a single starburst ($10^{10}\,M_{\odot}$ in stars), to roughly describe elliptical galaxies and globular clusters.

\section{Results} \label{3. Results}
\begin{figure}
   \centering
   \includegraphics[width=9cm,angle=0]{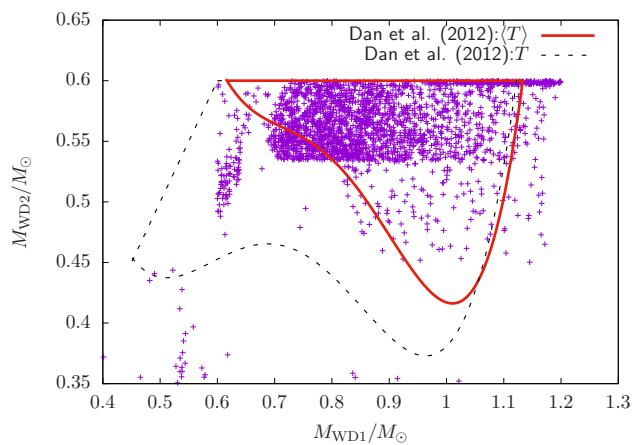}
   \caption{BPS results of CO WD$+$He-rich WD systems in the $M_{\rm WD1}-M_{\rm WD2}$ plane, where $M_{\rm WD1}$ and $M_{\rm WD2}$ are the merging mass of the CO WD and He-rich WD, respectively. Here, $\alpha_{\rm CE}$ is set to be 1. The contours with the solid and dashed line represent the region of the parameter space for producing SNe Ia given by Dan et al. (2012), in which the individual particle temperature $T$ and the SPH-smoothed temperature <$T$> are adopted, respectively.}
\end{figure}

\subsection{CO WD$+$He-rich WD systems}
\begin{table}
\begin{center}
 \caption{Birthrates and delay times of SNe Ia based on the merger scenario of a CO WD and a He-rich WD. Notes: $\nu$, DTDs, and $M_{\rm upper}$ are the birthrates, the delay time distributions, and the upper mass limit for He-rich WDs, respectively.}
   \begin{tabular}{cccccc}
\hline \hline
 Set & $\alpha_{\rm CE}\lambda$ & $M_{\rm upper}(M_\odot)$ & $\rm\nu (10^{\rm -4}{\rm yr}^{\rm -3})$ & $\rm DTDs ({\rm Myr})$\\
\hline
$1$ & $0.3$ & $0.6$ & $0.20$ & $450-8900$\\
$2$ & $0.5$ & $0.6$ & $0.40$ & $>220$\\
$3$ & $1.0$ & $0.6$ & $4.47$ & $>110$\\
$4$ & $0.3$ & $0.55$ & $0.09$ & $890-8900$\\
$5$ & $0.5$ & $0.55$ & $0.13$ & $>280$\\
$6$ & $1.0$ & $0.55$ & $1.34$ & $>180$\\
\hline \label{1}
\end{tabular}
\end{center}
\end{table}

In order to investigate the properties of SNe Ia from the merger scenario of a CO WD and a He-rich WD, we performed a series of Monte Carlo BPS simulations from the formation of primordial systems to the merging moment of double WDs (see Table\,1). According to the merger scenario of a CO WD and a He-rich WD, we found that the masses of the primordial primaries for producing SNe Ia are in the range of $2.4$$-$$6.1\,M_{\odot}$, the masses of the primordial secondaries range from $1.7$ to $6.0\,M_{\odot}$, and the primordial orbital periods are in the range of $230$$-$$6700\rm\,days$.

Figure\,2 shows the distribution of CO WD$+$He-rich WD systems in the $M_{\rm WD1}$$-$$M_{\rm WD2}$ plane, where $M_{\rm WD1}$ and $M_{\rm WD2}$ are the merging mass of the CO WD and He-rich WD, respectively. Here, $\alpha_{\rm CE}$ is set to be 1.0 and metallicity $Z$ is assumed to be 0.02. We note that the absolute number of CO WD$+$He-rich WD systems decreases and the relative numbers does not change significantly when we adopt a lower CE efficiency (e.g., $\alpha_{\rm CE}=0.3$ or 0.5). All these double WDs can merge in the Hubble time. In this figure, we only provide CO WD$+$He-rich WD systems with masses of He-rich WDs higher than $0.35\,M_{\odot}$, although many CO WD$+$He-rich WD systems have masses of He-rich WDs that are lower than $0.35\,M_{\odot}$. We also show the region of the parameter space for producing SN explosions obtained by Dan et al. (2012), in which the individual particle temperature $T$ (dashed line) and the SPH-smoothed temperature <$T$> (solid line) is used. This comparison is only based on the WD masses. In the present work, the region obtained via the SPH-smoothed temperature was adopted to calculate the properties of SNe Ia from the merging of CO WDs and He-rich WDs.

\subsection{Birthrates and delay times of SNe Ia}
\begin{figure}
   \centering
   \includegraphics[width=6.5cm,angle=270]{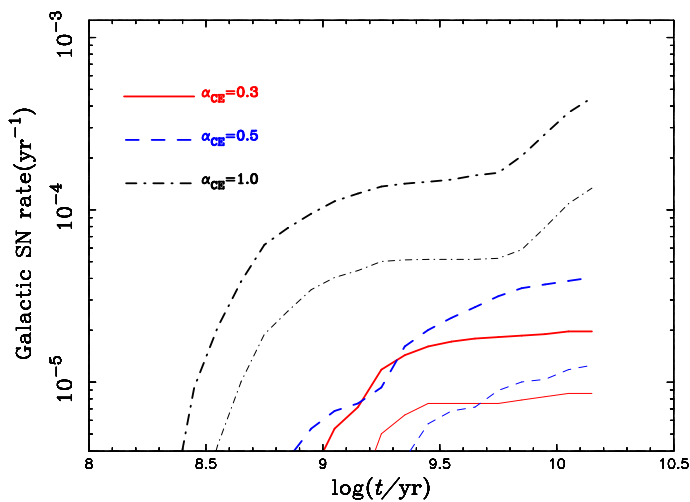}
   \caption{Evolution of Galactic SN Ia birthrates with time for a constant Population I SFR (${\rm SFR}=5\, M_{\odot}{\rm yr}^{-1}$) based on the merger scenario of a CO WD and a He-rich WD. The thick curves show the cases with an upper mass limit of He-rich WDs equal to $0.6\,M_{\odot}$, while the thin curves represent the cases with a mass limit equal to $0.55\,M_{\odot}$. The solid, dashed, and dash-dotted curves represent the cases with a CE ejection efficiency $\alpha_{\rm CE}=0.3$, 0.5, and 1.0, respectively.}
\end{figure}

Figure\,3 shows the evolution of the Galactic birthrate of SNe Ia as it changing with time by adopting a star formation rate of $5\,M_{\odot}yr^{\rm -1}$ and a metallicity of $Z=0.02$.
The theoretical Galactic birthrates of SNe Ia from the merger scenario of a CO WD and a He-rich WD are in the range of $\sim$$2.0\times10^{\rm -5}\,\rm yr^{\rm -1}$ to $\sim$$4.5\times10^{\rm -4}\,\rm yr^{\rm -1}$. Observations show that the Galactic SN Ia birthrate is $\sim$$3$$-$$4\times 10^{\rm -3}\,\rm yr^{\rm -1}$ (e.g., Cappellaro \& Turatto 1997). Hence, this scenario may contribute to $\sim$$0.5\%$$-$$15\%$ of all SNe Ia in the Galaxy. The theoretical birthrates of SNe Ia from the merger scenario of a CO WD and a He-rich WD are lower than observational results, which indicates that this merger scenario may only form a certain type of SNe Ia. We note that the SN Ia birthrate decreases for a larger $\alpha_{\rm CE}$. The reason
is that the CE is more difficult to eject and the ejection of the CE releases more orbital energy when a lower $\alpha_{\rm CE}$ is adopted, resulting in more mergers during the CE phase.

The delay-time distributions of SNe Ia can be used to examine progenitor models by comparing the theoretical with observational
predictions (e.g., Ruiter et al. 2009, 2011; Toonen et al. 2012). In Fig.\,4 we present the delay-time distributions with various values of $\alpha_{\rm CE}$ by assuming a single starburst with $10^{10}\,M_{\odot}$ in stars. The delay times of SNe Ia from the merger scenario of a CO WD and a He-rich WD range from $110\,\rm Myr$ to the Hubble time. This scenario mainly contributes to SNe Ia in intermediate and old populations. We note that the large end of $\log(t)$ for the case of $\alpha_{\rm CE}=0.3$ is a real cutoff based on our numerical simulations, while the cutoffs for other cases are artificial as the time has already reached the Hubble time.

\begin{figure}
   \centering
   \includegraphics[width=6.5cm,angle=270]{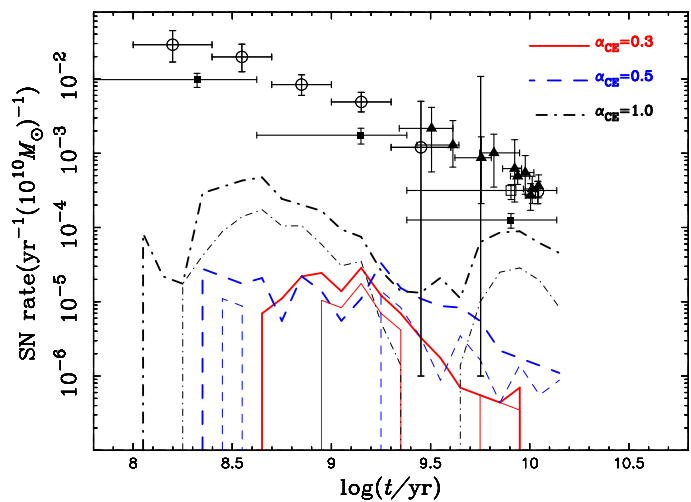}
   \caption{Similar to Fig.\,3, but for the delay-time distributions of SNe Ia. The open square is from Graur \& Maoz (2013), the filled triangles and squares are from Maoz et al. (2010, 2012), and the open circles are taken from Totani et al. (2008).}
\end{figure}

The upper mass limit for He-rich WDs is still uncertain. In the present work, we set the upper mass limit to be $0.6\,M_{\odot}$ (see also Dan et al. 2012, 2014; Yungelson \& Kuranov 2016). However, the upper mass limit for He-rich WDs is likely to be lower, for example, $0.55\,M_{\odot}$ (e.g., Han et al. 2000). In Figs. 3-4, the thin curves show the evolution of the Galactic birthrate and delay-time distributions of SNe Ia for the cases with the masses of He-rich WDs lower than $0.55\,M_{\odot}$. In this case, the theoretical Galactic birthrates of SNe Ia are from $\sim$$8.6\times10^{\rm -6}\,\rm yr^{\rm -1}$ to $\sim$$1.3\times10^{\rm -4}\,\rm yr^{\rm -1}$, contributing to $\sim$$0.2\%$$-$$4.3\%$ of all SNe Ia in the Galaxy. The delay times of SNe Ia from the merger scenario of a CO WD and a He-rich WD are in the range of $178\,\rm Myr$ to the Hubble time. It appears that the upper mass limit for He-rich WDs has a significant influence on the birthrates and delay times of SNe Ia from the merger scenario of a CO WD and a He-rich WD. Thus, more numerical simulations are needed to constrain the upper mass limit of He-rich WDs.

\subsection{Mass distributions of CO WD$+$He-rich WD}
There are several ongoing projects that are searching for double WDs as SN Ia progenitor candidates, e.g., ESO-VLT Supernova-Ia Progenitor Survey (e.g., Napiwotzki et al. 2004; Nelemans et al. 2005; Geier et al. 2007) and SWARMS survey (e.g., Badenes et al. 2009). According to our BPS calculations, we can present some properties of CO WD$+$He-rich WD systems for producing SNe Ia, which may be useful for searching progenitor candidates of SNe Ia.

In the double-detonation model, the exploding masses of the CO WDs play a key role in the production of $^{\rm56}\rm Ni$ and the brightness distribution of SNe Ia (e.g., Sim et al. 2010; Ruiter et al. 2013). In Fig.\,5 we plot the distribution of the final masses of CO WDs in CO WD$+$He-rich WD systems for producing SNe Ia with different $\alpha_{\rm CE}$. Here, we set the metallicity $Z=0.02$ and we assumed an ongoing constant star formation rate. According to our calculations, we found that CO WD$+$He-rich WD systems with high-mass CO WDs are mainly produced from Channels{\em } A and B in Fig.\,1.
This figure shows that a lower $\alpha_{\rm CE}$ forms CO WDs that are more massive.
For a lower $\alpha_{\rm CE}$, the formed double WDs originate from primordial binaries with wider separations, and those with smaller separations will merge during the CE phase and form new AGB stars, which therefore do not contribute to the double WD population that can produce SNe Ia via the CO WD$+$He-rich scenario.
It is obvious that there is a second peak around $1.0$$-$$1.1\,M_{\odot}$ only for the case of $\alpha_{\rm CE}=0.5$.
We note that if we do not consider the restriction of the contour presented in Fig. 2, the final mass distribution of CO WDs will also show double peaks for the cases of $\alpha_{\rm CE}=0.3$. However, the right-hand peak for the case with $\alpha_{\rm CE}=0.3$ is beyond the right-hand boundary of the contour shown in Fig.\,2. These double peaks are mainly produced from different formation channels in Fig.\,1; the left-hand peak is mainly from Channel{\em } A and the right-hand peak is mainly produced from Channel{\em } B. For the extreme case of $\alpha_{\rm CE}=1.0$, a large number of CO WD+He-rich WD systems that contribute to the left-hand peak are produced by Channel{\em } A, the number of which is so large that it engulfs the right-hand peak, resulting in a single-peaked distribution.
Ruiter et al. (2013) also obtained a double-peaked WD mass distribution, resulting from the mass-accretion of CO WD from a He-rich companion before merger, although this mass-accretion process is not relevant here (see also Liu et al. 2016).

Figure\,6 shows the distribution of the final masses of He-rich WDs in CO WD$+$He-rich WD systems. This figure shows that the masses of He-rich WDs are higher than $0.45\,M_{\odot}$. Since all He WDs have a mass lower than $0.46\,M_{\odot}$, hybrid HeCO WD may be the only option for WDs with masses slightly higher than $0.45\,M_{\odot}$ (e.g., Sweigart et al. 1990). Thus, we suggest that in order to produce SNe Ia, the companion of CO WD relies on a hybrid HeCO WD but not a He WD based on the merger scenario of a CO WD and a He-rich WD. In Fig.\,7 we display the distribution of the total masses of CO WD$+$He-rich WD systems for producing SNe Ia. The total masses are in the range of 1.2$-$$1.75\,M_{\odot}$. This distribution with $\alpha_{\rm CE}=0.5$ also has two peaks, which is caused by the double peaks of the masses of CO WDs. We note that the fractions of CO WD$+$He-rich WD systems with total masses higher than the Chandrasekhar limit are in the range of 45\%$-$86\% for different $\alpha_{\rm CE}$.

\begin{figure}
   \centering
   \includegraphics[width=10cm,angle=0]{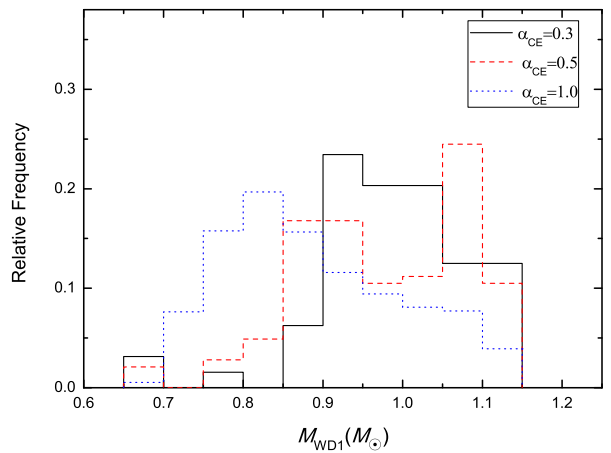}
   \caption{Mass distributions of CO WDs in the CO WD$+$He-rich WD systems with $\alpha_{\rm CE}=0.3, 0.5$ and 1.0. Here, every case is normalized to be 1.}
\end{figure}

\begin{figure}
   \centering
   \includegraphics[width=10cm,angle=0]{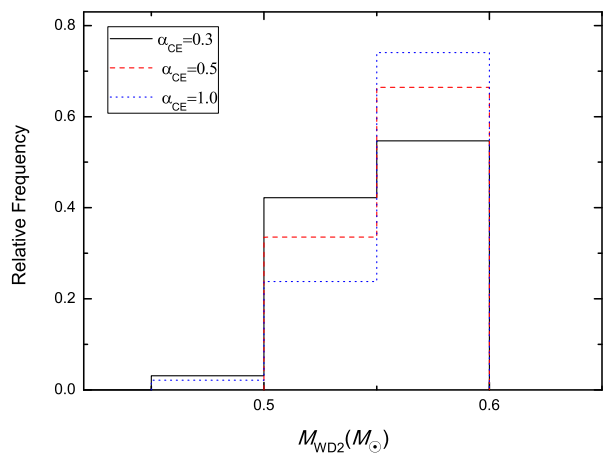}
   \caption{Similar to Fig.\,5, but for the mass distributions of He-rich WDs.}
\end{figure}

\begin{figure}
   \centering
   \includegraphics[width=10cm,angle=0]{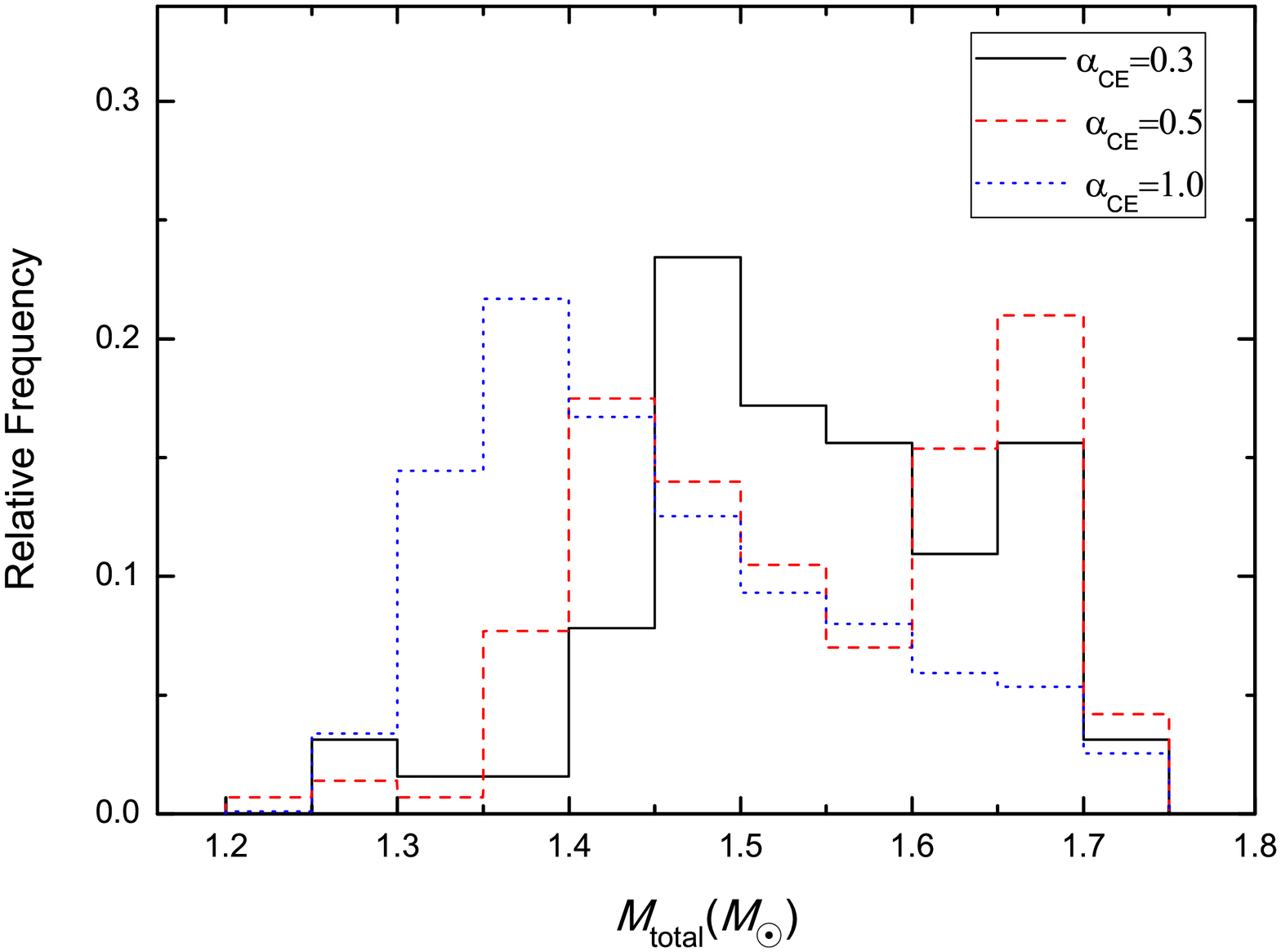}
       \caption{Similar to Fig.\,5, but for the distributions of total masses of double WDs.}
\end{figure}

\subsection{Influence of metallicities}
It has been suggested that there may be a correlation between SN Ia properties and metallicity in observations (e.g., Branch \& Bergh 1993; Hamuy et al. 1996; Wang et al. 1997; Cappellaro et al. 1997). Numerical calculations also show that the metallicity may influence the final amount of $^{\rm56}\rm Ni$ and the light curves of SNe Ia (e.g., Timmes et al. 2003; Travaglio et al. 2005; Podsiadlowski et al. 2006).
In Fig.\,8 we present the BPS results of SNe Ia with various metallicities (i.e., $Z=0.0001, 0.01, 0.02$, and 0.03), in which $\alpha_{\rm CE}$ is set to be 1.0. We employed the same IMF for all metallicities. This figure shows that the birthrates of SNe Ia range from $\sim$$4.1\times10^{\rm -4}\,\rm yr^{\rm -1}$ to $\sim$$6.6\times10^{\rm -4}\,\rm yr^{\rm -1}$ for different metallicities. The birthrates of SNe Ia are higher for the cases with low metallicities.
This is because the less massive primordial binaries that are not massive enough to produce SNe Ia in a higher-$Z$ model retain their mass because of the decreased wind in a lower-$Z$ model, and thus they contribute to the formation of CO WDs that are more massive. According to the initial mass function, there are more less massive primordial binaries than massive binaries.
We also note that when we evolve the same binaries that can form SNe Ia in the lower-$Z$ model, the cores of the secondaries are too low to produce SNe Ia in a higher-$Z$ model when the second CE is formed.

\begin{figure}
   \centering
   \includegraphics[width=6.5cm,angle=270]{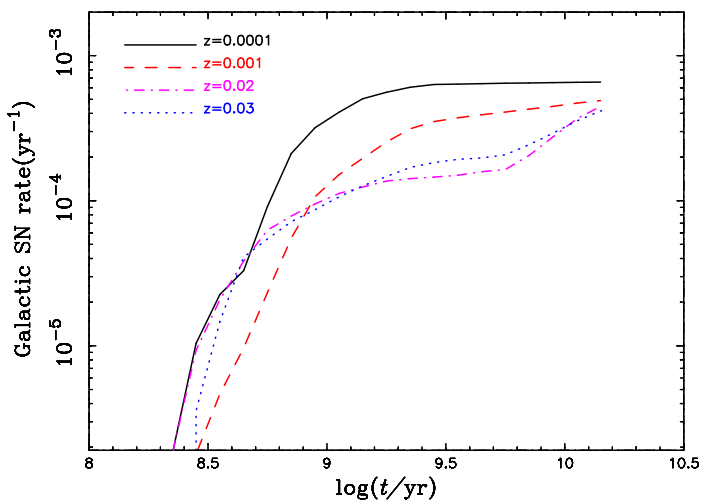}
   \caption{Similar to Fig.\,3, but for various metallicities. Here, $\alpha_{\rm CE}$ is set to be 1.0.}
\end{figure}

\section{Discussions and summary} \label{4. Discussion}
The peak luminosity of SN 1991bg-like events are fainter than the luminosity in normal SNe Ia, and their light curves decline faster (Filippenko et al. 1992; Leibundgut et al. 1993). The low peak luminosity of SN 1991bg-like events reveals a small amount of $^{\rm 56}\rm Ni$ produced during these events ($\sim$$0.07\,M_{\odot}$, see Mazzali et al. 1997). This type of events belongs to old populations (>1 Gyr, see Sullivan et al. 2006), and their birthrates account for about $16\pm7\%$ of all SNe Ia (Li et al. 2001, 2010). The subluminous property of SN 1991bg-like events is generally explained by the double-detonation model from a CO WD$+$He star system or the merger of a CO WD$+$He-rich WD system as the exploding CO WDs in the double-detonation model have sub-Chandrasekhar masses (Branch et al. 1995; Livio 1999, 2003). Crocker et al. (2016) recently suggested that the merging of CO WD and He WD systems may produce SN 1991bg-like events, and argued that this type of events might be the main source of Galactic positrons. In this work, we found that the merging of a CO and a He-rich WD may contribute to $\sim$0.5$-$15\% of all SNe Ia, and their delay times range from $\sim$$110\,\rm Myr$ to the Hubble time. This indicates that the merger scenario of a CO WD and a He-rich WD could match the birthrates and cover the delay times of SN 1991bg-like events.

The spectra and colors of SNe Ia from the double-detonation model simulated by Kromer et al. (2010) favor a hybrid HeCO companion star. The authors found that the light curves of SNe Ia from their simulations cover the range of brightness and the rise and decline times of SN Ia observations, while the colors and spectra are too red to match the observed ones since too much $^{\rm 44}\rm Ti$ and $^{\rm 48}\rm Cr$ is produced during the explosion of the CO core. Kromer et al. (2010) found that the theoretical colors and spectra of SNe Ia would match the observations if the He stars contain 34\% of $^{\rm 12}\rm C$ by mass. Interestingly, this component abundance is similar to the composition of hybrid HeCO WDs (e.g., Dan et al. 2014). According to the calculations of Han et al. (2000), the mass fractions of carbon in hybrid HeCO WDs are in the range of 20$-$95\% and increase with the WD masses, which are in the range of $0.32$$-$$0.6\,M_{\odot}$.

Many WD binary systems have been found in
observations, and some of them might be (or will evolve into) CO WD$+$HeCO hybrid WD systems. These double WDs include V458 Vulpeculae with $P_{\rm orb}$$\sim$$0.3\,\rm d$, $M_{\rm1}$$>$$1.0\,M_{\odot}$ , and $M_{\rm2}$$\sim$$0.6\,M_{\odot}$ (Rodr\'iguez-Gil et al. 2010) and SBS 1150$+$559A with $P_{\rm orb}$$\sim$$0.163\,\rm d$, $M_{\rm1}$$\sim$$0.86\,M_{\odot}$ , and $M_{\rm2}$$=$$0.54$$\pm$$0.02\,M_{\odot}$ (Tovmassian et al. 2010). Additionally, KPD 1930$+$2752 and CD$-$$30^{\circ}\, 11223$ are two WD$+$sdB binary systems that may eventually evolve into CO WD$+$HeCO hybrid WD systems. KPD 1930$+$2752 and CD$-$$30^{\circ}\, 11223$ have current parameters of ($M_{\rm WD}$, $M_{\rm sdB}$, $P_{\rm orb}$)$=$($\sim$$0.97\,M_{\odot}$, $\sim$$0.55\,M_{\odot}$, $2.283\,\rm h$) and ($\sim$$0.76\,M_{\odot}$, $\sim$$0.51\,M_{\odot}$, $1.2\,\rm h$), respectively (Geier et al. 2007; Geier et al. 2013). All of these binaries have a merging time shorter than the Hubble timescale. More observations of these systems are needed to constrain their properties.

However, the outcome of the merging of CO WD$+$He-rich WD systems may not be merely SNe Ia. In order to reproduce the observation properties of the faintest SNe Ia, Sim et al. (2010) suggested that the minimum mass of the exploding CO WD is $0.8\,M_{\rm \odot}$ (see also Ruiter et al. 2013). Our results show that the fraction of CO WD$+$He-rich WD systems with a CO WD mass lower than $0.8\,M_{\rm \odot}$ ranges from 0.5\% to 24\% based on different values of $\alpha_{\rm CE}$. The merging of these systems may only trigger the detonation of He-rich shell and the CO WD is likely to survive, which corresponds to an SN .Ia event (e.g., Bildsten et al. 2007; Piersanti et al. 2015). If the first detonation of the He-rich shell results in an off-center ignition of a CO WD, the CO WD will be converted into an ONe WD. If the ONe WD approaches the Chandrasekhar mass limit, it will eventually evolve into a neutron star initiated by the electrons capture of Ne and Mg (e.g., Nomoto 1984; Podsiadlowski et al. 2004; Kitaura et al. 2006). Perets et al. (2010) recently suggested that a binary with a primary WD and a He-rich secondary may also produce faint 2005E-like events (see also Meng \& Han 2015).

We here performed a series of Monte Carlo BPS simulations to study the birthrates and delay times of SNe Ia based on the merger scenario of a CO WD and a He-rich WD. We found that this scenario contributes to $\sim$0.5$-$15\% of all SNe Ia in our Galaxy, and the delay times from this scenario range from $110\,\rm Myr$ to the Hubble time. We note that this scenario can roughly reproduce the birthrates of SN 1991bg-like events, and it covers the range of their delay-time distributions.
The total masses of CO WD$+$He-rich WD systems required to produce SNe Ia are in the range of 1.2$-$$1.75\,M_{\odot}$. The fraction of the systems with a total mass higher than the Chandrasekhar limit is in the range of 45\%$-$86\% for the cases with different $\alpha_{\rm CE}$.
We also found that the companion of a CO WD relies on a hybrid HeCO WD based on the merger scenario of a CO WD and a He-rich WD.
In order to place further constraints on this merger scenario of SNe Ia, more observations and numerical simulations on this scenario are required.

\begin{acknowledgements}
We acknowledge useful comments and suggestions from the anonymous referee.
We also thank Xuefei Chen for her helpful discussions.
This work is supported by the National Basic Research Program of China (973 programme, 2014CB845700),
the National Natural Science Foundation of China (Nos 11673059, 11521303, 11390374  and 11573016),
the Chinese Academy of Sciences (Nos KJZD-EW-M06-01),
the Key Research Program of Frontier Science CAS (No QYZDB-SSW-SYS001),
the Youth Innovation Promotion Association CAS,
and the Natural Science Foundation of Yunnan Province (Nos 2013HB097 and 2013HA005).
\end{acknowledgements}

\end{document}